\documentstyle[12pt]{article}

\newcommand{\noi}{\noindent}
\newcommand{\be}{\begin{equation}}
\newcommand{\ee}{\end{equation}}
\newcommand{\bea}{\begin{eqnarray}}
\newcommand{\eea}{\end{eqnarray}}
\newcommand{\ba}{\begin{array}}
\newcommand{\ea}{\end{array}}

\begin{document}
\date{}

\centerline{\bf QUANTUM COSMOLOGY FOR A QUADRATIC THEORY OF GRAVITY}
\bigskip
\centerline{Luis O. Pimentel$^1$ and Octavio Obreg\'on$^{1,2}$}
\centerline{$^1$Departamento de F\'{\i}sica,}
\centerline{Universidad Aut\'onoma Metropolitana,}
\centerline{Apartado Postal 55-534,}
\centerline{CP 09340 M\'exico D.F.,MEXICO.}
\centerline {$^2$Instituto de F\'{\i}sica de la }
\centerline {Universidad de Guanajuato}
\centerline {Apartado Postal E-143,}
\centerline {C. P. 37150 Le\'on, Guanajuato, MEXICO.}
\bigskip
\vfill
\eject
\centerline{\bf ABSTRACT}
For pure fourth order (${\cal{L}} \propto R^2$) quantum cosmology the
Wheeler-DeWitt equation is solved exactly for the closed
homogeneous and isotropic model. It is shown that by imposing as
boundary condition that $\Psi = 0$ at the origin of the universe
the wave functions behave as suggested by Vilenkin.


\vfill
\eject
\parskip=8pt
\bigskip

\bigskip
\section{Introduction}

Non-linear modifications of the Eintein-Hilbert action, have a long history
[1],
and they are of interest, among others, for the following reasons: first,
because the
effective gravitational action predicted by closed bosonic, heterotic and
supersymmetric strings give the usual Einstein term plus a correction quadratic
 in the curvatures [2]. Second, these theories can be renormilized when
quantized [3]. Third, pure gravity inflationary models emerge on adding an
$R^2$ term to the usual gravitational Lagrangian [4]. On the other hand, some
non-linear Lagrangians can be chosen with the property that the field equations
for the metric are second order, these are the so-called Lovelock actions [5]
which can be regarded as formed by the dimensional continuation of the Euler
characteristics of lower dimensions [6,7,8]. These Gauss-Bonnet terms
seem to be of importance for the quantization of these theories.

In this paper we consider a toy quantum model for the theories
mentioned. The action will be constructed only with the square of
the Ricci scalar and the corresponding Wheeler-De Witt equation for
the closed isotropic cosmological model will be solved by
separation of variables.
The Vilenkin conditions will result after choosing the vanishing
of the wave function at the beginning of the universe, as first
proposed by DeWitt [9]. Previous work considering  non-linear gravitational
Lagrangians has been made by
Hawking and Luttrell [10], Boulware [11], Buchbinder and Lyanhovich
[12], and Vilenkin [13] among others.

\section {Wheeler DeWitt Equation}
The $R+\gamma R^2$ theory of gravity, at the classical level is related
to Einstein's theory of gravity with a scalar field as the source
of gravity by a conformal transformation [14], a similar relation
exists for the scalar-tensor
theories of gravitation [15]. This relation has been exploited by Kasper
[16]
to obtain approximate solutions to the pure quantum cosmology of
fourth order.

The Theory studied by Kasper for the spatially closed Robertson-Walker
metric,

\be
ds^2=-c^2dt^2+a^2(t)\left [ {dr^2\over 1- r^2}+
r^2(d\theta^2+\sin^2\theta d\phi^2) \right ],
\ee
\noi where $a(t)$ is the expansion factor of the universe, leads to the
following form of the Wheeler~-De Witt equation in fourth order quantum
cosmology:

\be
[{ {\partial^2}\over {\partial \beta ^2}}- {{\partial^2}\over {\partial
\alpha
\partial\beta}} - 12 e^{6\alpha + 3\beta}+ 144e^{4\alpha +
2\beta}]  \; {\cal \psi} (\alpha,\beta)= 0.
\label{eq:edif1}
\ee

\noi where $ \alpha=\ln a,  \beta=\ln R$ and $R$ is the scalar
curvature; therefore this model is
restricted to $R >0$, since the above equation was obtained using the signature
-+++,
classical anti deSitter space is excluded. Notice that the lapse function in
that paper
is taken equal to 1 in contrast to the paper of Hawking and
Luttrell where it is proportional to the scale factor. Therefore
the time coordinates are different in the two cases.

In order to solve the above equation it is useful to introduce the
following change of independent variables,

\be
 x=e^{2 \alpha + \beta}=a^2 R, \; y= \alpha + \beta=\ln(aR),
\label{eq:change}
\ee

\noi the new variables extend over the half plane $x>0$ (since $R>0 $).
The resulting differential equation is now

\be
[x { {\partial^2}\over {\partial x^2}}+ {{\partial^2}\over {\partial
x
\partial y}} +{{\partial}\over {\partial x}}+ 12 x^2- 144x]  \; {\cal \psi}
(x,y)= 0.
\label{eq:edif2}
\ee

\noi We look for a solution by separation of variables,

\be
\psi (x,y) = F(x) G(y).
\label {eq:sepa}
\ee
After substitution into the differential equation we obtain

\be
G'=\nu G,
\ee

\be
x F'' + (\nu +1) F' +[ -144 x + 12 x^2]F=0,
\label{eq:eqx}
\ee

\noi where $\nu$ is the separation constant.

The solution for G is trivial, $G= G_0 e^{\nu y}$. The solution for  F
is a little more complicated, if $ \nu \ne \pm 1 $ it can be obtained as
an infinite series. But, if $\nu=\pm 1$ it is possible to obtain closed
form solutions.
\subsection{$\nu =-1$ Solution}
In this case Eq.(\ref{eq:eqx}) reduces to

\be
 F''+[-144  +12 x]F=0
\label{eq:eqminus}
\ee

\noi This equation can be transformed into a Bessel equation
of order 1/3 in the variable $u=-12^{4/3}+12^{1/3}x, F=P(u)$,

\be
P'' +u P=0.
\ee

\noi  The solution can be written as follows

\be
F = F_1 Ai (-u) + F_2 Bi (-u)
\ee

\noi here Ai and Bi are the Airy functions and $F_1, F_2$ are constants that
will be fixed by the boundary conditions. The complete state function of the
universe is

\be
\psi(x,y)=e^{-y} [c_1 Ai(-u) + c_2 Bi(-u) ]
\label{eq:solminus}
\ee

\noi Taking the limit of this solution when x is large results into the
WKB solution obtained in reference [16]

\be
F \to  (-u)^{-1/4} e^{\pm {2\over 3} u^{3/2}}= (- (12^{1/3}x)^{-1/4}
e^{\pm{2\over 3} (-12^{1/3}x )^{3/2} }
\ee

The particular solution of this section is also important when looking at
the asymptotic solution ($x \to \infty$) in the case of arbitrary value
for $\nu$ as will be shown in the next section.

\subsection{Boundary condition and WKB approximation}

If we restore the original variables $a$ and $R$ in
Eq.(\ref{eq:solminus})
we have

\be
\psi(a,R)=\frac  {c_1 Ai[-12^{1/3}(a^2 R-12)] + c_2 Bi[-12^{1/3}(a^2 R-
12)]}{aR},
\label{eq:solar}
\ee

\noindent at $a=0$ the values of the Airy functions in the numerator of
the wave function of the universe are finite
in contrast with the denominator that vanishes. Therefore the only
sensible thing to do is choosing the constants $c_1$ and $c_2$ so that
the whole numerator vanishes at $a=0$, and that corresponds to the
following choice,

\be
c_2=\frac{-c_1 Ai[12^{4/3}]}{Bi[12^{4/3}]}.
\label{eq:con}
\ee
\noi With this choice the numerator in Eq.( \ref{eq:solar})
dominates the behaviour at the beginning of the universe and the
wave function vanishes at $a=0$.

We recall here that the vanishing of the wave function of the
universe at $a=0$ is the boundary condition which Bryce DeWitt [9]
argued must be imposed on it, because
 it has the effect of keeping the wavepackets away from the
singularity. On the other hand if we consider Eq.(\ref{eq:eqminus})
 and apply the WKB method to it, taking into account the right
boundary conditions for the case where the potential goes to
infinity, that is the case here because our coordinate x is
defined only for positive values,  we obtain,

\be
\psi(x)_{WKB} = \frac  { Sinh [\int_{0}^{x} \sqrt{(12(12-x))} dx
]}{[ 12(12-x) ]^{1/4}} ,
\label{eq:solwkb}
\ee

Now the WKB solution vanishes at $a=0$ (x=0).
The behaviour of the wave function that we have obtained is similar to the
hydrogenic wave functions when the angular momentum $l \ne 0$ in the sense
that they vanish at the origin of the radial coordinate.

\subsection{$\nu =1 $Solution}

If in Eq.(\ref{eq:eqx}) we eliminate the first derivative term, to
put the equation into its normal form by means of the transformation

\be
F(x)= x^{-(\nu +1)/2} g(x).
\ee

\noi the differential equation is now

\be
g''+[-12(12-x)+ \frac{1 - \nu^2}{4x^2}]g=0,
\label{eq:eqnormal}
\ee

\noindent and we notice that for $\nu = 1$ the differential equation for g(x)
is the
same as Eq.(\ref{eq:eqminus}). Therefore the complete wave
function of the universe is

\be
\psi(x,y)=e^{+y}\frac{ [c_1 Ai(-u) + c_2 Bi(-u) ]}{x}.
\ee

\noi Here we also take as the boundary condition the vanishing of
the wave function at
$a=0$, that means that the constants $c_i$ are given by Eq. (
\ref{eq:con}).

\bigskip

\section{Series Solutions}

Before looking at the series
solution to Eq. (6) for $\nu \ne \pm 1$, it is
useful to consider the asymptotic
limit $x \to \infty$ . For that purpose, we look once more at the normal
form of the
differential equation \ref{eq:eqnormal}

\be
u''+[-12(12-x)+ \frac{1 - \nu^2}{4x^2}]u=0
\ee

\noi and it is clear that in the asymptotic region ($x \to
\infty$) this equation reduces to
the equation solved exactly in the previous section.

In the case $\nu \ne \pm 1$ we can use Frobenius method assuming a
solution for Eq.(6) of the form

\be
F(x)= |x|^{\alpha}\sum_{n=0}^{\infty}{a_n x^n}.
\ee

\noi The indicial equation is given by

\be
\alpha(\alpha+\nu)=0,
\ee
\noi and the solutions are

\be
\alpha_1=0, \alpha_2=-\nu.
\ee

\noi The recurrence relations are

\be
a_n(n+\alpha-\alpha_1)(n+\alpha-\alpha_2)= 144 a_{n-2}-12 a_{n-3}
\ee

\noi after substituting the solutions to the indicial equation we
have

\be
a_n(n+\alpha)(n+\alpha+\nu)= 144 a_{n-2}-12 a_{n-3}.
\ee

\noi From the above equations the series for both solutions is
completely determined in case $\nu $ is not an integer; if
that is the case the second solution can be obtained by the
standard procedure used in Frobenius method.

\bigskip
\section{Final remarks}

The simplified model used in this paper has allowed us to exactly
solve, for the first time, the quantum cosmology of fourth order
for a closed Friedmann model. The analogous problem for spatially
flat model was solved by Reuter [17].
Actions close related to Chern-Simmons are of particular interest
at the present time; we consider our present work as an attempt in
the program of studying the quantum cosmology of these more general
Lagrangians of topological origin.
\bigskip
\section{ Acknowledgment}
\bigskip
This work was partially supported by CONACYT GRANTS 1861-E9212,
1683-E9209 and F246-E9207.

\end{document}